\documentclass{iau-JDSS}
\usepackage{graphicx}

\title[Detecting individual gravity modes in the Sun] 
{Detecting individual gravity modes in the Sun: Chimera or reality?}

\author[Garc\'\i a et al.]   
{Rafael A. Garc\'\i a
  \thanks{In collaboration with: J. Ballot, A. Eff-Darwich, R. Garrido, A. Jim\'enez, S. Mathis, S. Mathur, A. Moya, P.L. Pall\'e, C. R\'egulo, D. Salabert, K. Sato, J.C. Su\'arez and S. Turck-Chi\`eze
}
  }

\affiliation{Laboratoire AIM, CEA/DSM -- CNRS - Universit\'e Paris Diderot -- IRFU/SAp, \\91191 Gif-sur-Yvette Cedex, France\break email: rafael.garcia@cea.fr\\[\affilskip]}

\pubyear{2009}
\volume{Volume 15}  
\pagerange{119--126}
\date{?? and in revised form ??}
\jname{Highlights of Astronomy, Volume 14}
\editors{Ian F Corbett, ed.}
\begin{document}

\maketitle

\begin{abstract}
Over the past 15 years, our knowledge of the interior of the Sun has tremendously progressed by the use of helioseismic measurements. However, to go further in our understanding of the solar core, we need to measure gravity (g) modes. Thanks to the high quality of the Doppler-velocity signal measured by GOLF/SoHO, it has been possible to unveil the signature of the asymptotic properties of the solar g modes, thus obtaining a hint of the rotation rate in the core (Garc\'\i a \textit{et al.} 2007, 2008a). However, the quest for the detection of individual g modes is not yet over. In this work, we apply the latest theoretical developments to guide our research using GOLF velocity time series. 
In contrary to what was thought till now, we are maybe starting to identify individual low-frequency g modes...
\end{abstract}

\section{Observations and analysis}

Gravity modes are very sensitive to the structure (e.g. \cite{Bas09,Gar08c}) and the dynamics  (e.g. \cite{Mat08}) of the radiative zone and, in particular, to the inner core of the Sun. There have been many attempts to look for them without so far an undisputed detection of such modes (\cite{App09}), although some interesting peaks and patterns have been detected with GOLF and VIRGO (e.g. \cite{Stc04,Jim09}) with a high confidence level.

A 4500-day GOLF time series (\cite{Gab95}) starting on April 11, 1996 and calibrated into velocity (\cite{Gar05})  has been used to compute a single, full resolution power spectrum density (PSD) in spite of the different sensitivity to the visible solar disk between the blue- and the red-wing GOLF measurements (see for further details \cite{Gar98,Ulr00}). To increase the signal-to-noise ratio, we have also smoothed the PSD  with a 41-nHz boxcar function as it is commonly done in asteroseismology when the signal is weak (e.g. \cite{Mic08}). 

\section{Discussion}
\begin{figure}[!htb]
\begin{center}
 \includegraphics[angle=90,width=0.8\textwidth]{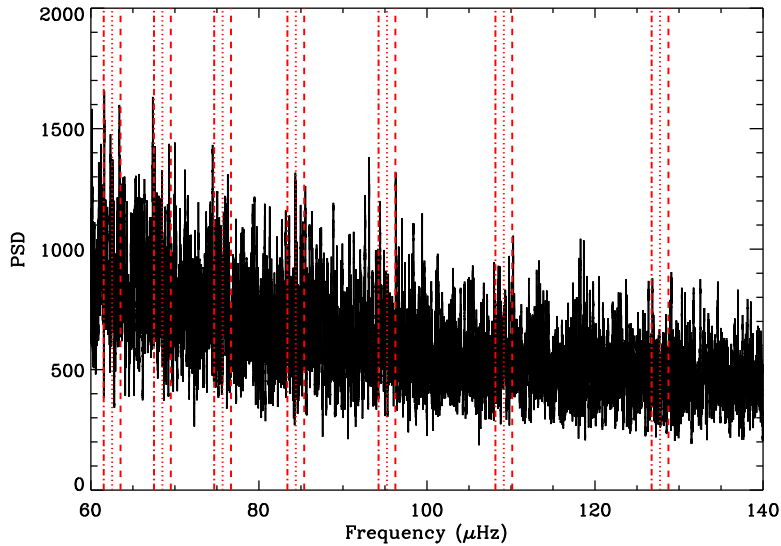} 
 \caption{GOLF PSD. The dotted vertical lines are the central frequencies of the dipole g modes computed by the Saclay seismic model. The vertical dashed and dot-dashed lines are the rotational split components with a rotational splitting 4.5 times larger than in the radiative zone above 0.2~$R_{\odot}$.}
   \label{fig1}
\end{center}
\end{figure}

Several of the highest peaks between 60 and 140 $\mu$Hz are located around the theoretical frequencies of the dipole modes obtained by the Saclay seismic model (e.g. \cite{Mat07}). Moreover, varying the splitting of the modes from 1 to 5 times the rotation rate of the radiative region, $\Omega_{\mathrm{rad}}$,  we notice that there is a quasi complete sequence of peaks matching the model when the splitting of these modes is around 4.5 $\Omega_{\mathrm{rad}}$ (see Fig.\ref{fig1}). This could be the first time that individual g modes are identified in the Sun. For example, the candidate mode $\ell$=1, n=$-4$ has an amplitude of 1.8 $\pm$ 0.4 mm/s (with a signal-to-noise ratio of $\sim$ 4) which is close to the latest theoretical predictions (\cite{Bel09}).

The splittings can be estimated for each individual g-mode candidate. However, it is very difficult to obtain an accurate inference of the rotational profile of the core (\cite{Gar08b,Mat09}). The rotation rate in the core could be up to $\sim$7 $\Omega_{\mathrm{rad}}$, since $\sim$60$\%$ of the rotational kernel of these g modes is located in the inner core.


\end{document}